\documentclass{PoS}
\usepackage{url}
\title{INTEGRAL IBIS/ISGRI energy calibration in OSA 10}

\ShortTitle{INTEGRAL IBIS/ISGRI energy calibration in OSA 10} 

\author{\speaker{I.~Caballero}$^a$, 
  J.~A.~Zurita~Heras$^b$,
  F.~Mattana$^b$,
  S.~Soldi$^{a,b}$,
  P.~Laurent$^b$,
  F.~Lebrun$^b$,
  L.~Natalucci$^c$,
  M.~Fiocchi$^c$,
  C.~Ferrigno$^d$,
  R.~Rohlfs$^d$\\
    \llap{$^a$}Laboratoire AIM, CEA/IRFU, CNRS/INSU, Universit\'e Paris Diderot, CEA DSM/IRFU/SAp, 91191 Gif-sur-Yvette, France\\
 \llap{$^b$}APC, Astroparticule et Cosmologie, UMR 7164, Universit\'e Paris Diderot, Sorbonne
Paris-Cit\'e, CNRS/IN2P3, Observatoire de Paris, 10, rue A.~Domon et L.~Duquet, 75205 Cedex 13, France\\
 \llap{$^c$}INAF-Istituto di Astrofisica e Planetologia Spaziali, Via del Fosso del Cavaliere 100, I-00133 Roma, Italy \\
  \llap{$^d$} ISDC Data Center for Astrophysics, University of Geneva, chemin d'\'Ecogia, 16, 1290 Versoix, Switzerland\\
  E-mail: \email{isabel.caballero@cea.fr}}
\abstract{We present the new energy calibration of the ISGRI detector onboard 
\textsl{INTEGRAL}, that has been implemented in the Offline Scientific Analysis (OSA) version 10.  
With the previous OSA~9 version, a clear departure from stability 
of both W and $^{22}$Na background lines was observed after MJD$\sim$54307 (revolution $\sim$583). 
To solve this problem, the energy correction in OSA 10 uses: 
1) a new description for the gain depending on the time and the pulse rise time, 2) an improved 
temperature correction per module, and 3) a varying shape of the low threshold, corrected for the change in energy resolution. 
With OSA 10, both background lines show a remarkably stable behavior with a relative energy 
variation below 1\% around the nominal position ($>$6\% in OSA 9), 
and the energy reconstruction at low energies is more stable compared to previous OSA versions. 
We extracted Crab light curves with ISGRI in different energy bands using all available data since the beginning 
of the mission, and found a very good agreement with the currently operational hard X-ray instruments 
\textsl{Swift}/BAT and \textsl{Fermi}/GBM.
}

\FullConference{An INTEGRAL view of the high-energy sky (the first 10 years)"
9th INTEGRAL Workshop and celebration of the 10th anniversary of the launch,\\
		October 15-19, 2012\\
		Bibliotheque Nationale de France, Paris, France}

\begin{document}

\section{IBIS/ISGRI energy calibration}
\subsection{Energy calibration with OSA~9} 
\begin{figure}[!h]
\centering
\includegraphics[angle=0,width=0.49\textwidth]{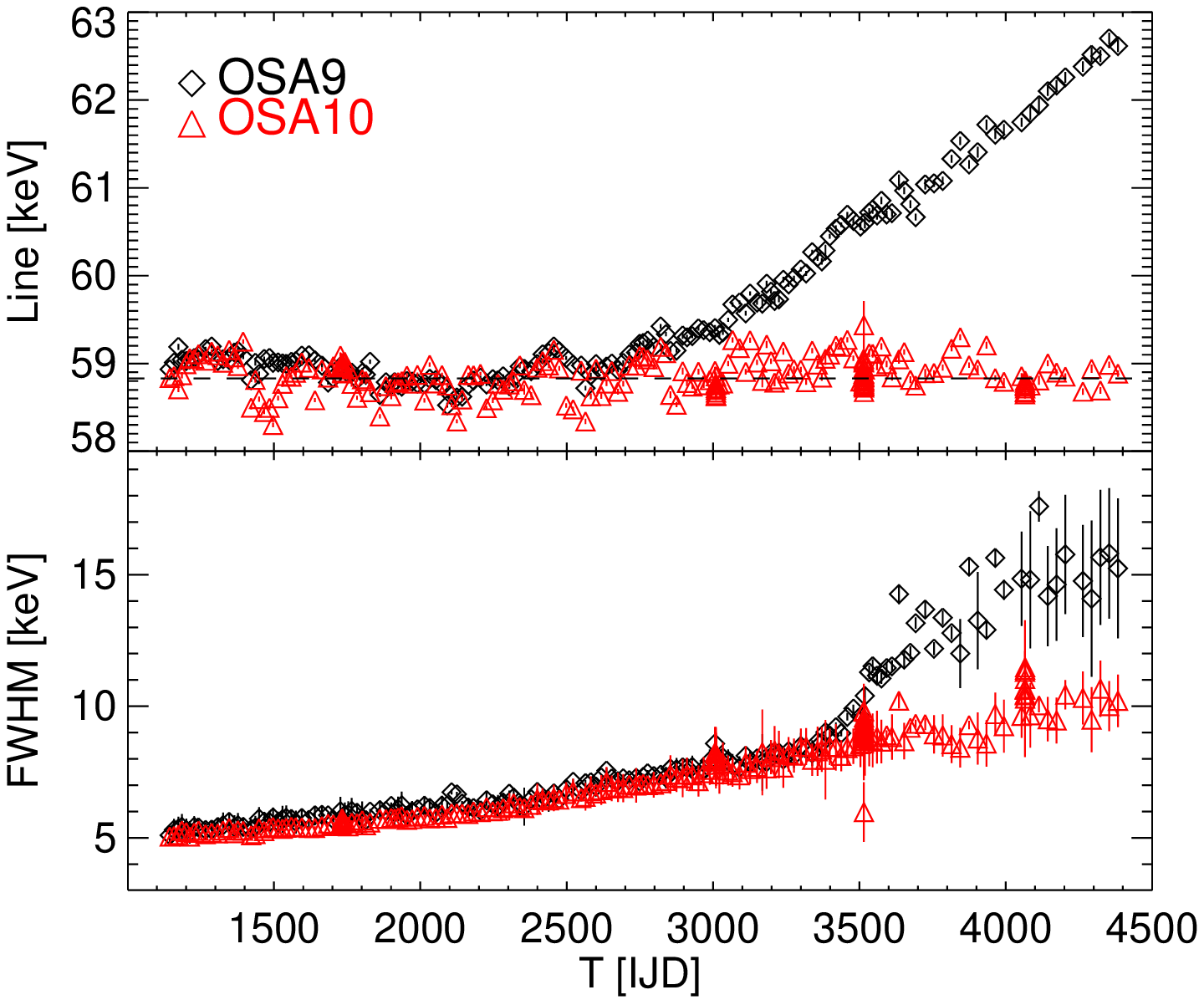}
\includegraphics[angle=0,width=0.49\textwidth]{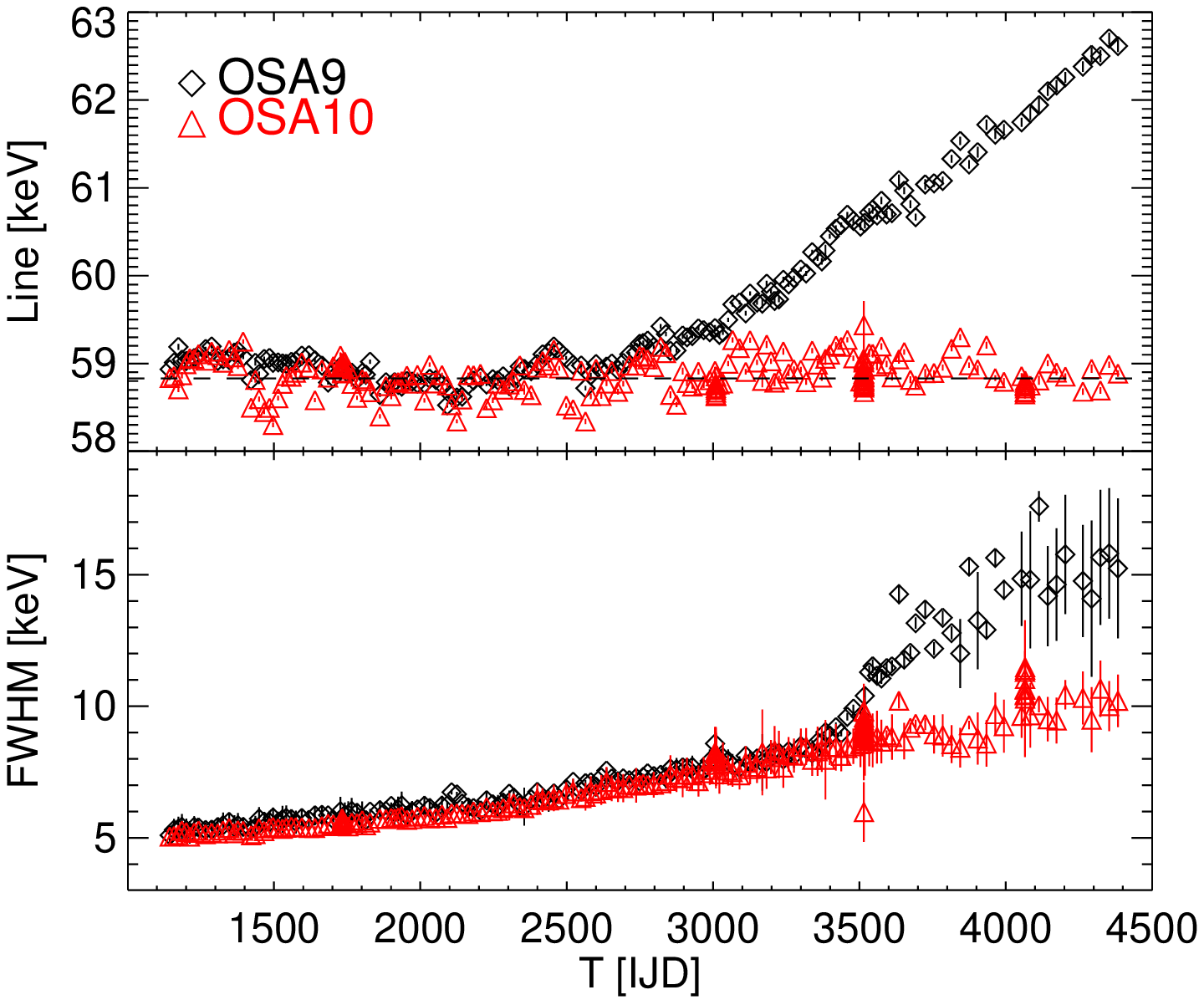}
\vspace{-0.1cm}
\caption{\label{fig:W_Na}\footnotesize{\textsl{Left:} evolution of the W fluorescence line position (top) and 
FWHM (bottom) obtained with OSA~9 (black) and OSA~10 (red). \textsl{Right:} Evolution of the $^{22}$Na line position 
(top) and FWHM (bottom) obtained with OSA 9 (black) and OSA 10 (red). The dashed horizontal lines in the upper 
panels represent the nominal positions of the W  and $^{22}$Na lines.}}
\end{figure}

ISGRI \cite{lebrun03} is the low energy detector of the IBIS imager \cite{ubertini03} on-board \textsl{INTEGRAL} \cite{winkler03}, and is 
made of 128x128  cadmium telluride (CdTe) pixels grouped in 8 Modular Detector Units (MDU).  
The scientific data analysis is performed with the Offline Scientific Analysis software package
OSA, delivered by the ISDC\footnote{ISDC Data Centre for Astrophysics, \url{http://www.isdc.unige.ch/}}.  
The ISGRI spectral gain decreases with time. In the OSA~9 version, 
the description of the gain drift was based on the Radiation Environment Monitor (IREM) counters \cite{hajdas03} 
integrated over time, to take into account the solar flares. 
The gain is followed using the radioactive sodium ($^{22}$Na) and tungsten (W) fluorescence lines 
 located at 511 and 58.8297 keV\footnote{
This energy is the mean 
obtained between the K$\alpha_{1}$ (59.3 keV)  and K$\alpha_{2}$ (57.98 keV) lines \cite{deslattes03}. 
Using the ratio I($\alpha_{2}$)/I($\alpha_{1}$)=0.57 from \cite{kasagi95}, the mean energy 
is  58.8297\,keV.}, respectively. As shown in Fig.~\ref{fig:W_Na}, the energy reconstruction used in 
OSA~9 is not valid since IJD\footnote{\textsl{INTEGRAL} Julian Date, IJD=MJD-51544.}$\sim$2763 (revolution number $\sim$583). 
The position of the background fluorescence lines shows a gradual increase with time. 

\subsection{New energy calibration with OSA 10} 

\subsubsection{Temperature correction}
The temperature and voltage dependence of the gains and offsets of the events rise time and pulse height was evaluated
on ground and in flight \cite{terrier03}. In previous OSA versions, the temperature offset of each module with regard
to the average was assumed to be constant, i.e., a stable thermal map. This assumption turned out to be wrong.
A more accurate temperature correction has been introduced in OSA~10, in which the temperatures 
from the ISGRI thermal probes in each detector's module are used, instead of assuming a constant $\Delta T$.  
The mean temperature of the ISGRI modules varies by $\sim$15--20$^{\circ}$C, and the maximum temperature 
difference between the modules is about $3\,^{\circ}$C.

\subsubsection{Gain drift}

In OSA 10, the gain drift is measured using the W and $^{22}$Na background lines between revolutions 42 and 1106. 
{To increase the statistics, particularly important in the  $^{22}$Na region, 
the background lines are measured in bins of 15 revolutions. 
The pulse height gain and offset are then described as a function of the pulse rise time and the time. 
Fig.~\ref{fig:LT} (left) shows the pulse height gain for different rise time intervals as a function of time. 
For energies below $\sim$50\,keV, charge loss is negligible. We assume that the pulse height offset 
is constant and that the gain evolution does not depend on the rise time, and the gain 
is modeled with a linear function of time. For energies above $\sim$50\,keV,  the pulse height gain and offset 
are modeled both as a function of the rise time and time. After the correction, 
both background lines show a remarkably stable behavior, as shown in Fig. 1, with a relative energy variation below 1\% 
around the nominal position (>6\% in OSA~9). The FWHM of the W line increases by a factor ~2 between revolutions 39 
and 1142 (instead of ~3 with OSA~9), indicating a better energy reconstruction. 

\section{Low threshold correction} 
The low threshold (LT) position is corrected with the new energy calibration (Fig.~\ref{fig:LT}, right). 
The LT is stable in channel units. 
Since OSA~10, the LT shape follows the evolution of the spectral resolution, instead of being fixed at the 
W line resolution at the beginning of the mission as in previous OSA versions. 
Therefore, its resolution corresponds to the W line resolution, and also evolves with time. The accuracy achieved is around 1\%. 
The jumps in Fig.~\ref{fig:LT} (right) correspond to 
different uploaded LT settings over the whole mission duration. After ten years in orbit, the lower threshold is still below 23\,keV.

\begin{figure}[h]
\centering
\includegraphics[angle=90,width=0.49\textwidth]{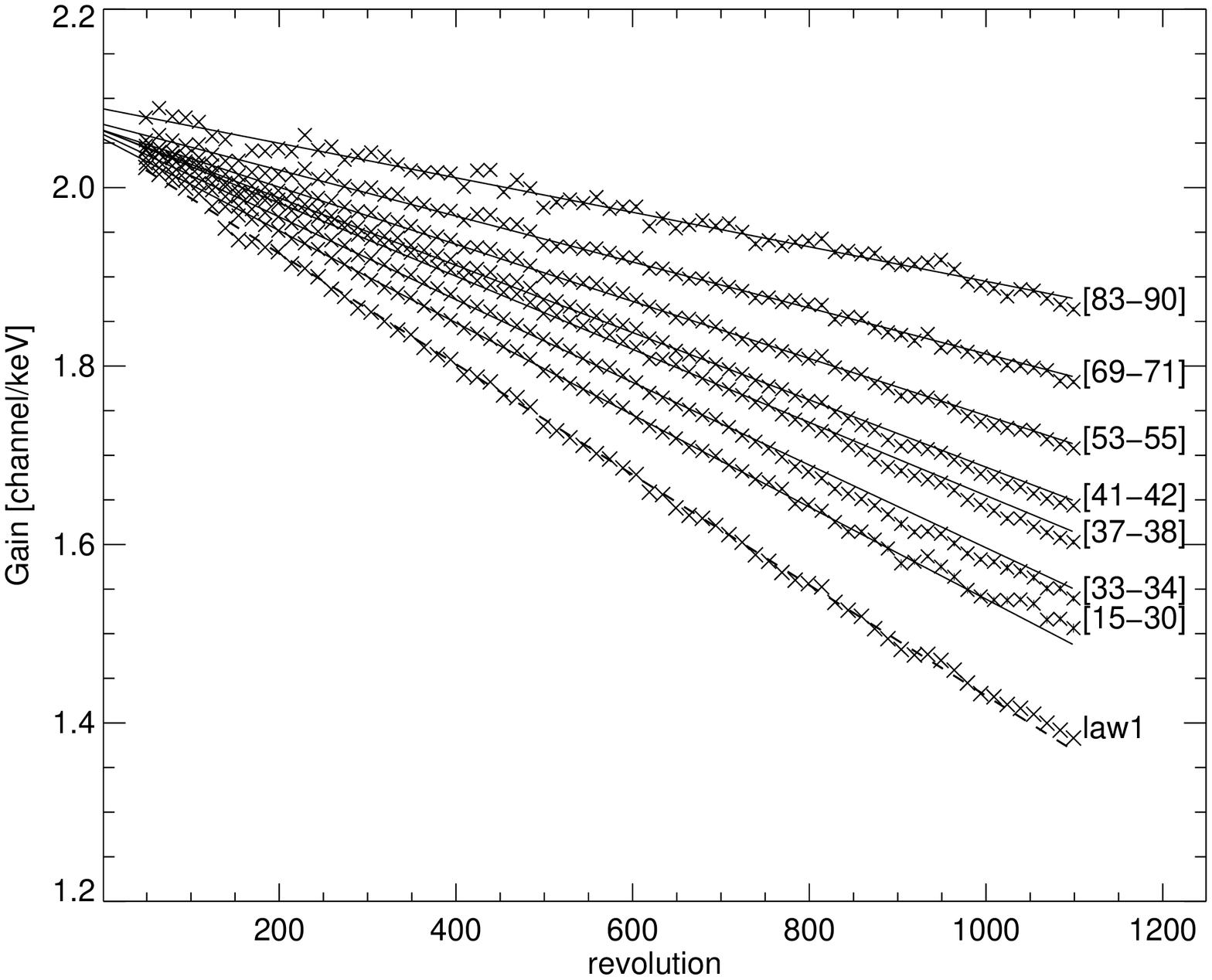} 
\includegraphics[width=0.45\textwidth]{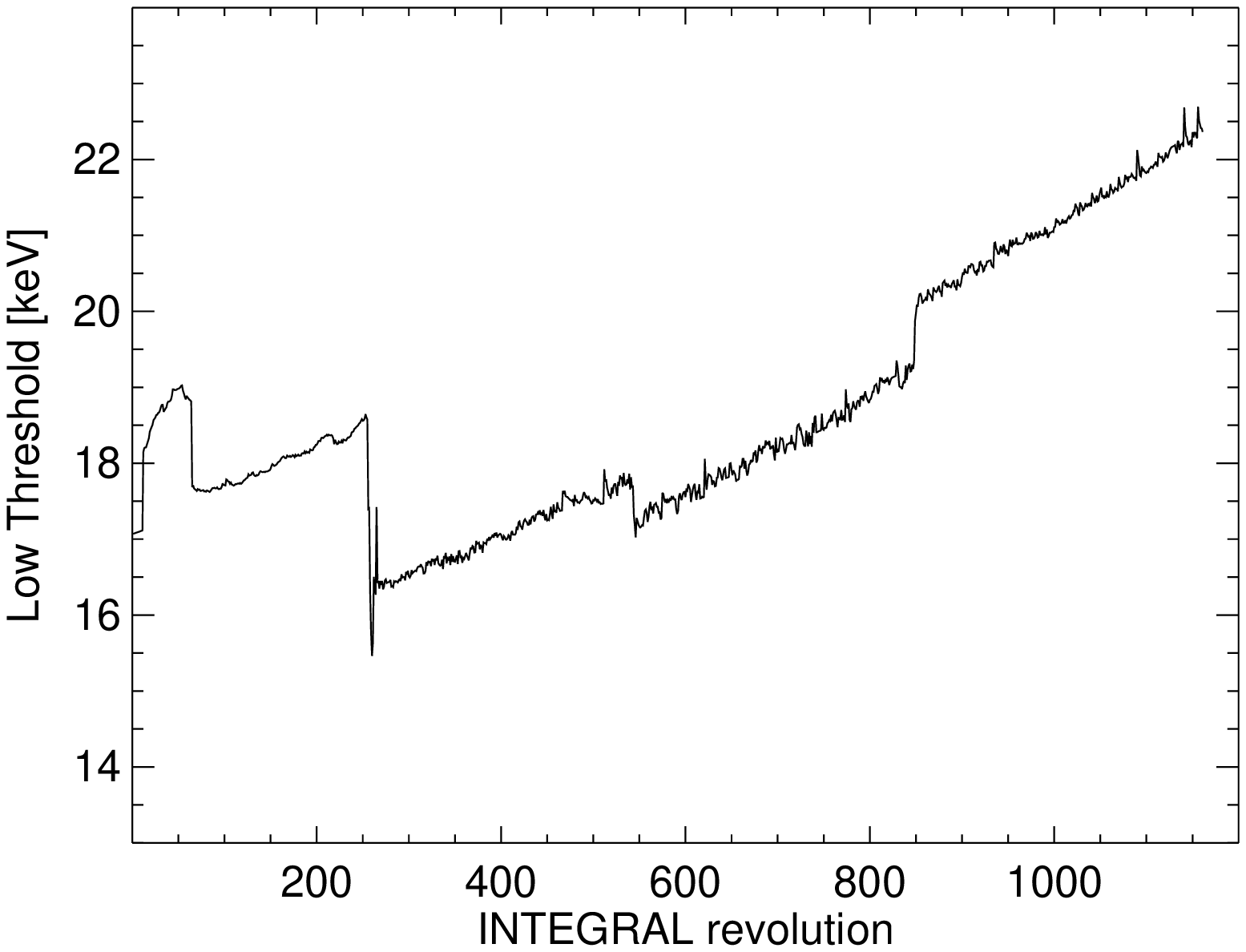} 
\vspace{-0.1cm}
\caption{\label{fig:LT}\footnotesize{\textsl{Left:} Pulse height gain evolution for several rise time intervals (labeled in the figure) as a function of the revolution number. 
 The linear fits used to describe the gain for the different rise time intervals are overplotted. The evolution 
of the gain for energies below $\sim$50\,keV, for which a constant offset is assumed (see text), is also plotted, labeled as ``law 1''.  \textsl{Right:} Low threshold position with OSA~10.}}
\end{figure}

\section{Spectral analysis} 
Crab and background spectra extracted with OSA~9 and OSA~10 for a sample of revolutions are shown in Figs.~\ref{fig:crab} and 
~\ref{fig:back}. By comparing the left and right panels of Fig.~\ref{fig:crab}, 
the OSA~10 correction results in a more stable spectrum along the mission (especially at high energy) with
respect to OSA~9.

A set of ancillary response files (ARFs) has been produced for different epochs using Crab observations. As an example, 
a fit of the Crab spectrum from revolution 839 extracted with OSA~10 is shown in Fig.~\ref{fig:spec}. The parameters obtained from the spectral 
fit to a broken power law, reported in Table~\ref{tab:tab}, are in good agreement with the expected values (see, e.g., \cite{jourdain09}). 

\begin{figure}
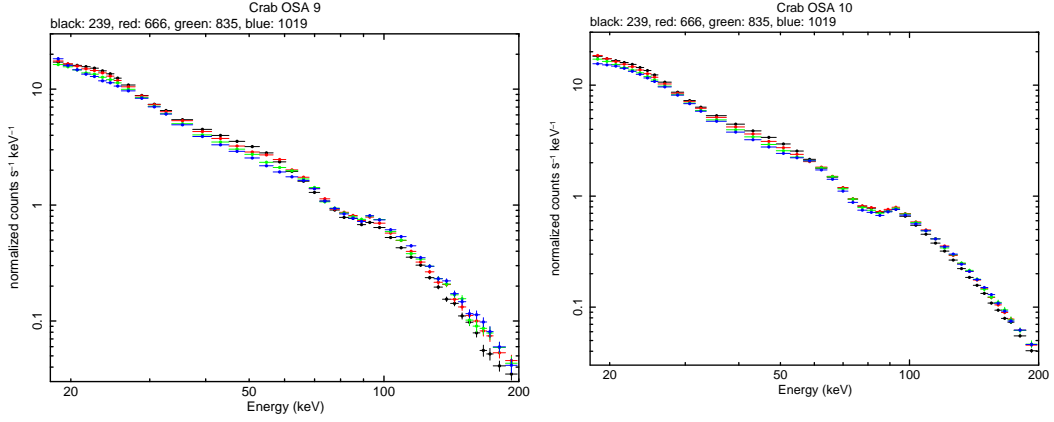

\hfill
\includegraphics[angle=270,width=0.47\textwidth]{fig5.eps}
\includegraphics[angle=270,width=0.45\textwidth]{fig6.eps}
\vspace{-0.1cm}
\caption{\label{fig:crab}\footnotesize{Crab spectra for a sample of revolutions (239, 666, 835, and 1019) extracted with OSA~9 (\textsl{left}) 
and OSA~10 (\textsl{right}).}}
\end{figure}
\begin{figure} 
\hfill
\includegraphics[angle=270,width=0.47\textwidth]{fig7.eps}
\includegraphics[angle=270,width=0.45\textwidth]{fig8.eps}
\vspace{-0.1cm}
\caption{\label{fig:back}\footnotesize{Background spectra for a sample of revolutions (239, 666, 835, and 1019) extracted with OSA 9 (\textsl{left}) 
and OSA 10 (\textsl{right}).}}
\end{figure}

\begin{figure}[h]
\centering
\includegraphics[angle=270,width=0.5\textwidth]{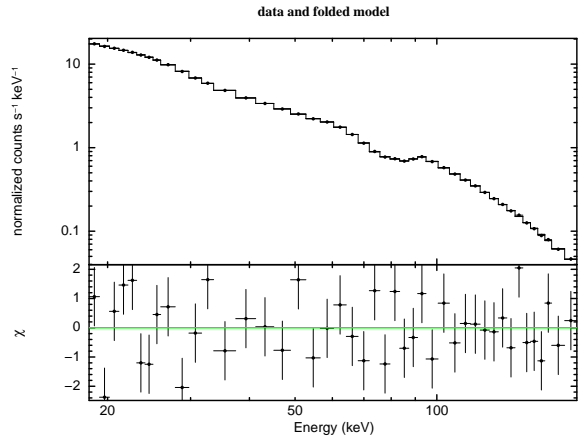}
\caption{\label{fig:spec}\footnotesize{ISGRI Crab spectrum for revolution 839 (top) and residuals of best fit model (bottom) extracted 
with OSA~10.}}
\end{figure}

\begin{table}
\caption{Best fit parameters for the ISGRI Crab observation in revolution 839. 0.3\% systematic errors are included. The quoted errors 
are at 90\% confidence level.} \label{tab:tab}
\begin{tabular}{|c|c|c|c|c|}\hline
  $\Gamma_{1}$       & $\Gamma_{2}$     &$E_{\mathrm{break}}$ [keV]&norm &$\chi^{2}_{\mathrm{red}}$/d.o.f.\\
& & & [photons\,keV$^{-1}$\,cm$^{-2}$\,s$^{-1}$ @ 1\,keV ]& \\\hline
$2.070\pm0.003$  & $2.24\pm{0.03}$ &$94^{+6}_{-5}$&$8.6\pm{0.1}$&1.15/40\\\hline
\end{tabular}
\end{table}

\section{Cross-calibration} 

Using the results of the imaging extraction, we built Crab light curves in the three energy bands 
25--50, 50--100, and 100--200\,keV. Figure~\ref{fig:crab_lc_all} shows the comparison between the 
ISGRI count rate and that of other currently operational hard X-ray instruments, \textsl{Swift}/BAT 
\cite{barthelmy05} and \textsl{Fermi}/GBM \cite{meegan09}. 
For each instrument, the light curve has been renormalized to the average count rate measured during the period MJD=[54690, 54790],
as in \cite{wilson11}. Note that the \textsl{Fermi}/GBM light curve in the highest energy band refers to the range 100--300\,keV. 
The \textsl{Fermi}/GBM light curves are from the GBM Occultation Project\footnote{\url{http://heastro.phys.lsu.edu/gbm/}}. 
For the \textsl{Swift}/BAT instrument, two different sets of light curves are reported, the 15--50\,keV one being provided by the 
\textsl{Swift}/BAT Hard X-ray Transient Monitor pages\footnote{\url{http://swift.gsfc.nasa.gov/docs/swift/results/transients/}}, 
and the 25--50, 50--100, and 100--200\,keV ones obtained from the Swift/BAT 58-months Hard X-ray survey\footnote{\url{http://swift.gsfc.nasa.gov/docs/swift/results/bs58mon/index.php}}. 
In general, there is a very good agreement between the light curves measured with the different instruments. 
Small differences are observed in the 25-50 keV band for 
MJD<54000 between ISGRI and BAT, and at E>100keV at MJD>55600 between ISGRI and GBM.

\begin{figure}
\centering
\includegraphics[angle=0,width=0.6\textwidth]{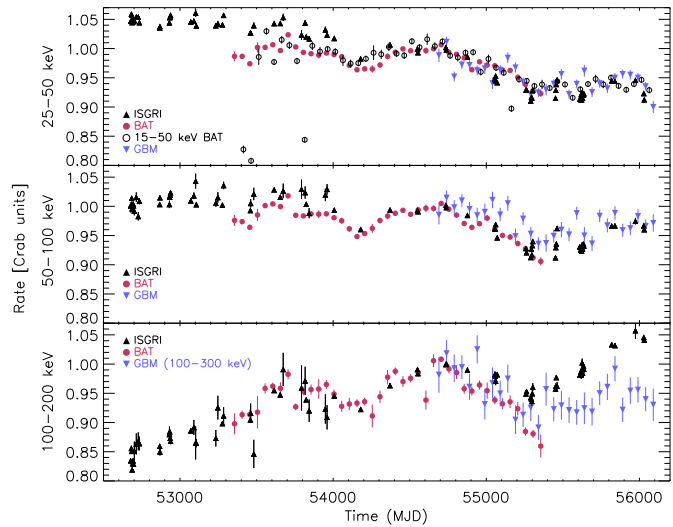}
\vspace{-0.1cm}
\caption{\label{fig:crab_lc_all}\footnotesize{\textsl{INTEGRAL}/ISGRI, \textsl{Swift}/BAT, and \textsl{Fermi}/GBM Crab light curves over the period of the 
\textsl{INTEGRAL} observations. Each light curve has been renormalized to its average value measured during the period MJD=[54690, 54790].}}
\end{figure}

\section{Conclusions}
A new energy correction has been implemented in OSA~10. The new calibration significantly improves the 
energy reconstruction. It includes a new description of the events gain and offset as a 
function of time and the events rise time, a more accurate temperature correction per ISGRI module, and a 
varying shape of the low threshold, corrected for the degradation of the spectral resolution. 
The background lines positions are remarkably stable. Very good agreement is obtained between the 
ISGRI long term Crab light curves and those obtained by other currently 
operational hard X-ray observatories, \textsl{Swift}/BAT and \textsl{Fermi}/GBM. 

The limitations of the current energy calibration and known issues that the user should be aware of are 
kept up to date in the IBIS Analysis User Manual, in the section ``Known Limitations'', available at the ISDC.

\section*{Acknowledgments} 
The present work is partly based on observations with INTEGRAL, an ESA
project with instruments and science data center (ISDC) funded by ESA
members states (especially the PI countries: Denmark, France, Germany,
Italy, Switzerland, Spain, Czech Republic and Poland, and with the
participation of Russia and the USA). ISGRI has been realized by
CEA-Saclay/DAPNIA with the support of the French Space Agency CNES. IC, JZH, and 
SS acknowledge financial support from CNES. FM acknowledges financial support from 
CNES and P\^ole Emploi. 

\end{document}